\documentclass[conference]{IEEEtran}
\IEEEoverridecommandlockouts
\usepackage{cite}
\usepackage{eso-pic}
\usepackage{amsmath,amssymb,amsfonts}
\usepackage{algorithmic}
\usepackage{graphicx}
\usepackage{textcomp}
\usepackage{xcolor}
\usepackage{textgreek}
\usepackage{microtype}
\def\BibTeX{{\rm B\kern-.05em{\sc i\kern-.025em b}\kern-.08em
    T\kern-.1667em\lower.7ex\hbox{E}\kern-.125emX}}
\begin{document}
\bstctlcite{IEEEexample:BSTcontrol}
\title{
Evaluation of EEG Foundation Models for Event-Based Burst-Suppression Detection in ICU}

\author{
\IEEEauthorblockN{
Elisa Vasta\IEEEauthorrefmark{1}\IEEEauthorrefmark{2},
Thorir Mar Ingolfsson\IEEEauthorrefmark{2},
Andrea Cossettini\IEEEauthorrefmark{2},\\
Luca Benini\IEEEauthorrefmark{2}\IEEEauthorrefmark{3},
Tilman Beck\IEEEauthorrefmark{1},
Emanuela Keller\IEEEauthorrefmark{1},
Una Pale\IEEEauthorrefmark{1}
}

\IEEEauthorblockA{
\IEEEauthorrefmark{1}Neurocritical Care Unit, Institute for Intensive Care Medicine and Department
of Neurosurgery,\\
University Hospital Z{\"u}rich and University of Z{\"u}rich, Z{\"u}rich, Switzerland
}
\IEEEauthorblockA{\IEEEauthorrefmark{2}Integrated Systems Laboratory, ETH Z{\"u}rich, Z{\"u}rich, Switzerland}
\IEEEauthorblockA{\IEEEauthorrefmark{3}DEI, University of Bologna, Bologna, Italy}
\thanks{Corresponding email: elisa.vasta@uzh.ch. Code available upon publication.}
}
\maketitle
\AddToShipoutPictureFG*{%
\AtPageLowerLeft{%
\put(\LenToUnit{0.75in},\LenToUnit{0.45in}){%
\parbox{\dimexpr\paperwidth-1.5in\relax}{%
\centering
\resizebox{\linewidth}{!}{%
\textcopyright{} This work has been submitted to the IEEE for possible publication. Copyright may be transferred without notice, after which this version may no longer be accessible.
}%
}%
}%
}%
}

\vspace{-1cm}
\begin{abstract}
Burst suppression (BS) is a clinically relevant electroencephalographic (EEG) pattern used to monitor sedation depth and brain activity in critically ill patients, particularly during induced coma in Intensive Care Units (ICUs). Automatic burst detection remains challenging because BS patterns vary substantially between patients and annotated datasets are scarce. 
Recently, EEG Foundation Models (FMs) have shown promise across several downstream EEG applications, but their usefulness for BS detection remains unexplored. We present the first study to evaluate EEG FMs for burst detection in reduced-montage ICU EEG without patient-specific calibration. We compare REVE-base, LUNA-large and LuMamba-Tiny with an adaptive thresholding baseline and a task-specific EEGNet baseline. Additionally, we complement conventional EEG window-based classification with event-based burst detection evaluation. This helps assessing clinically whether burst episodes are correctly detected, reducing the impact of expected annotation variability. The best model, REVE-base, achieved the highest event-based F1-score ($0.868 \pm 0.167$) and reduced burst-per-minute error by $52.1\%$ and $36.2\%$ compared to EEGNet and adaptive thresholding respectively, supporting FMs for scalable EEG monitoring in ICU.
Ablation experiments showed that full fine-tuning was the most effective adaptation strategy with respect to frozen-backbone training, two-step fine-tuning, and LoRA-based adaptation, improving event-based F1-score over frozen-backbone training by up to $+0.102$ for LUNA-large. With reduced labeled datasets, pretrained REVE-base outperformed random initialization by $+0.723$ event-based F1 points at 25\% of the cohort, demonstrating the benefit of pretraining FM representations when adapted to burst detection with limited labeled data.
\end{abstract}


\section{Introduction}
In the intensive care unit (ICU), routine sedation monitoring often relies on intermittent behavioral scales \cite{devlinClinicalPracticeGuidelines2018}, which become less informative during pharmacologically induced coma because patients are intentionally unresponsive. In this setting, electroencephalography (EEG) is needed to monitor brain activity, where burst suppression (BS) serves as an EEG-based marker of sedation depth and targeted as a therapeutic endpoint in refractory status epilepticus or refractory intracranial hypertension \cite{hermanConsensusStatementContinuous2015} \cite{cottenceauUseBispectralIndex2008}. 
According to the American Clinical Neurophysiology Society (ACNS) \cite{hirschAmericanClinicalNeurophysiology2021a}, BS is a continuous, quasiperiodic EEG pattern characterized by alternating bursts, high-voltage activity lasting between 0.5 and 30 s, and suppressions, isoelectric activity below 10 µV that occupies 50–99\% of the EEG recording.
Despite its clinical importance, BS monitoring in routine ICU practice remains largely dependent on intermittent visual EEG inspection and manual burst counting by clinicians or nurses, which are prone to inter-rater variability \cite{brandonwestoverRealtimeSegmentationBurst2013a}. Commercial EEG-derived indices such as the bispectral index (BIS) may support sedation monitoring \cite{devlinClinicalPracticeGuidelines2018}, but their role in ICU remains less established, and they do not directly quantify burst or suppression events. 

Automatic burst detection could therefore support more objective and continuous EEG monitoring, by enabling the estimation of measures such as bursts per minute (BPM) and burst-suppression ratio (BSR) that are clinically used to guide sedative titration \cite{cottenceauUseBispectralIndex2008} \cite{maElectroencephalographicBurstSuppressionPerioperative2022}. Reliable automatic BS detection remains challenging because annotated datasets are very scarce, ICU artifacts can contaminate EEG signal quality, and BS patterns vary across patients due to pathology, sedation protocols, or antiseizure medication \cite{loboDoesElectroencephalographicBurst2021a}.
Previous automatic BS detection methods have used handcrafted features with classical machine learning \cite{lofhedeClassificationBurstSuppression2008}\cite{alarcon-bragaDetectingDepthSedation2026}\cite{orgucTimeFrequencyDomainClassifier2025}, recurrent neural network \cite{alarcon-bragaDetectingDepthSedation2026}, fixed or adaptive thresholding \cite{brandonwestoverRealtimeSegmentationBurst2013a}, state-space estimation \cite{chemaliBurstSuppressionProbability2013}, and clustering methods \cite{baldassanoIRISModularPlatform2020}\cite{narulaDetectionEEGBurstsuppression2021a}\cite{furbassMonitoringBurstSuppression2016}.
Although these approaches demonstrate the feasibility of automatic BS detection, many remain sensitive to artifacts and signal variability, or rely on patient-specific calibration, which limits deployment in ICU settings, where models should generalize across patients and recording conditions.
This motivates the evaluation of generalized models such as EEG foundation models (FMs), which may provide transferable representations under limited annotated data. However, their usefulness for BS detection remains unclear.

To the best of our knowledge, this is the first study to investigate EEG FMs for BS detection, a clinically relevant yet underrepresented downstream task. The main contributions are:
\begin{itemize}
\item We provide the first evaluation of pretrained EEG FMs for generalized burst detection in six-channel reduced-montage ICU EEG, using a leave-one-subject-out (LOO) setting without patient-specific calibration.
\item We introduce a clinically oriented evaluation framework that assesses automatic burst detection at the event level and quantifies BPM estimation error.
\item We demonstrate that large fine-tuned EEG FMs outperform a task-specific EEGNet baseline and the standard recursive variance thresholding (RVT) method. The best model, REVE-base, achieved higher event-based F1-score ($0.868 \pm 0.167$) than EEGNet ($0.720 \pm  0.245$) and RVT ($0.812 \pm  0.233$), while reducing BPM error by \(52.1\%\) and \(36.2\%\) compared to EEGNet and RVT respectively.
\item We investigate the role of adaptation and pretraining in transferring EEG FMs to BS detection. Full fine-tuning improved event-based F1-score over frozen-backbone training by up to \(+0.102\) for LUNA-large, whereas pretrained initialization was particularly beneficial under limited labeled data, improving REVE-base event-based F1-score by \(+0.723\) at 25\% of labeled data compared with random initialization.

\end{itemize}

\section{Methods}

\subsection{Clinical datasets, preprocessing and annotations}
We used long-term ICU EEG recordings from 25 patients, each approximately one hour in duration, acquired during deep sedation in the Neurocritical Care Unit of the University Hospital Zurich (USZ) and previously described by Narula et al. \cite{narulaDetectionEEGBurstsuppression2021a}. Burst centers were manually annotated by two experts and converted into fixed 3.5 s burst intervals. Recordings were acquired at 200, 256, or 500 Hz using a reduced frontal-temporal montage. Six bipolar channels were extracted following \cite{narulaDetectionEEGBurstsuppression2021a}: Fp1-F7, F7-T3, T3-T5, Fp2-F8, F8-T4, and T4-T6. Signals were band-pass filtered between 1 and 40 Hz, notch-filtered at 50 Hz, and resampled to 200 Hz. The preprocessed EEG was segmented into 2s windows with a 1s step. Bursts were considered the positive class, and windows were labeled as burst when more than 50\% of their duration overlapped with a burst interval. 

\subsection{Models}
We compared EEG FMs with task-specific baselines for binary BS classification. We included LUNA-large \cite{donerLUNAEfficientTopologyAgnostic2025}, REVE-base \cite{ouahidiREVEFoundationModel}, and LuMamba \cite{broustailLuMambaLatentUnified2026} as FMs, using the original pretrained checkpoints released by the authors. REVE uses versatile positional embeddings for arbitrary electrode arrangements and signal lengths, LUNA maps variable EEG montages into a topology-agnostic latent space using learned-query cross-attention, and LuMamba combines latent channel unification with state-space temporal modeling for computational efficiency. These models were selected because they represent distinct montage-agnostic architectures. Bipolar channel locations not available in the original pretrained configurations were approximated as the midpoint of the corresponding electrode positions.
For each FM, we used the pretrained encoder as backbone and replaced the original classification head with a common lightweight binary classifier. Encoder outputs were mean-pooled across tokens and passed through layer normalization, dropout, and a one-hidden-layer MLP with hidden dimension 128 and dropout 0.1. This ensured a consistent comparison across models and adaptation strategies, with final model sizes of 69.2M parameters for REVE-base, 40.5M for LUNA-large, and 4.1M for LuMamba-Tiny.

As baseline, we implemented the standard method for BS monitoring, recursive-variance-thresholding, adapting the method from Westover et al. \cite{brandonwestoverRealtimeSegmentationBurst2013a}. The frontal EEG signal was obtained by averaging the FP1-F7 and FP2-F8 bipolar channels, and the forgetting factor was fixed to 0.9534, as reported for 200 Hz recordings. For each LOO fold, the classification threshold was optimized only on training and validation subjects using balanced error, avoiding patient-specific calibration and improbable all-burst or all-suppression predictions.
As an additional baseline, we trained EEGNet\cite{lawhernEEGNetCompactConvolutional2018}, a compact task-specific neural network representing an intermediate-complexity baseline between RVT and EEG foundation models. We used the EEGNet-8,2 configuration adapted to 2s, six-channel EEG windows sampled at 200 Hz.

\subsection{Adaptation strategies}
We evaluated four adaptation strategies for each EEG FM.
In full fine-tuning, all encoder parameters and the classification head were updated. In frozen-backbone training, the pretrained encoder was kept fixed and only the binary classification head was trained, assessing the transferability of pretrained representations.
In LoRA-based fine-tuning, low-rank trainable adapters were inserted into selected layers, while the remaining pretrained weights were frozen, enabling parameter-efficient task adaptation \cite{huLoRALowRankAdaptation2021a}. In two-step fine-tuning, the model was first trained with a frozen backbone for 5 epochs and then fully fine-tuned, aiming to stabilize adaptation before updating the encoder \cite{ouahidiREVEFoundationModel}.
To further assess the contribution of pretraining under limited annotated BS data, we compared pretrained and random initialization FMs using 25\%, 50\%, 75\%, and 100\% of training and validation data from each patient under the same downstream protocol. 

\subsection{Downstream training configuration}
Training hyperparameters were initialized from the fine-tuning protocol in \cite{donerLUNAEfficientTopologyAgnostic2025} and adapted to our downstream task. In particular, we performed a learning rate (LR) ablation and selected a conservative LR to stabilize the full fine-tuning of pretrained encoders on the limited annotated dataset.
All models were trained or fine-tuned using AdamW optimizer with LR of 2·10$^{-5}$, AdamW coefficients $\beta_1 = 0.9$ and $\beta_2 = 0.999$, and a weight decay of 0.05. We used a cosine LR schedule with 5 warm-up epochs, an initial warm-up LR of 2.5·10$^{-7}$, and a LR of 2.5·10$^{-6}$. Unless otherwise specified, models were trained for up to 30 epochs with a batch size of 256, and early stopping was applied based on the validation loss with a patience of 10 epochs. Validation was performed once per epoch. Training was performed using 32-bit precision on GPU. 
For fine-tuning strategies that updated the encoder, we applied layer-wise learning-rate decay with a factor of 0.75 to enable task adaptation while limiting disruption of pretrained representations.

\subsection{Performance evaluation}
Models were evaluated using a LOO strategy, where one patient was held out for testing and the remaining patients were used for training and validation. This setting is clinically relevant as it assesses whether the model can generalize to unseen ICU patients without patient-specific retraining or calibration. Predictions from overlapping 2s windows with a 1s stride were projected onto a 1s temporal grid. Each 1s epoch was labeled as burst if at least one of the two overlapping windows covering that epoch were predicted as burst.
Then, performance was assessed in a typical machine learning fashion, at the 1s window-sequence level using F1-score ('window-based F1'). To account for the clinical relevance of burst detection, we follow the event-based evaluation approach used for epileptic seizure detection \cite{danSzCORESeizureCommunity2025}: windows predicted as burst were grouped into burst events and compared with annotated burst intervals. Following clinical definition of bursts, event matching used a 0.5s tolerance on onset and offset; events longer than 30 s were split into multiple shorter events, and events separated by less than 1s were merged. 
Unlike window-based F1, which evaluates how well the model classifies each window, event-based F1 assesses whether clinically meaningful burst episodes are correctly detected. Finally, BPM was estimated from the predicted burst events and compared with the annotated BPM using the mean absolute error (MAE).

\section{Results and Discussion}
\begin{table}[!t] 
\caption{Model performance as mean $\pm$ standard deviation across LOO folds. 
Bold values indicate the best mean performance.}
\label{tab:usz_results}
\centering
\scriptsize
\renewcommand{\arraystretch}{1.2}
\setlength{\tabcolsep}{2pt}

\begin{tabular}{|p{0.25\columnwidth}|c|c|c|}
\hline
\textbf{Model} 
& \multicolumn{3}{c|}{\textbf{USZ}} \\
\cline{2-4}

& \textbf{\textit{Window-based F1}} 
& \textbf{\textit{Event-based F1}} 
& \textbf{\textit{Burst-per-minute MAE}} \\
\hline

EEGNet baseline
& $0.572 \pm 0.232$ 
& $0.720 \pm 0.245$ 
& $0.936 \pm 0.645$ \\
\hline

RVT baseline
& $0.508 \pm 0.225$ 
& $0.812 \pm 0.233$ 
& $0.702 \pm 0.369$ \\
\hline

LUNA-Large 40.5M
& $0.718 \pm 0.129$ 
& $0.838 \pm 0.120$ 
& $0.689 \pm 0.547$ \\
\hline

REVE-Base 69.2M
& $\mathbf{0.726 \pm 0.176}$ 
& $\mathbf{0.868 \pm 0.167}$ 
& $\mathbf{0.448 \pm 0.284}$ \\
\hline

LuMamba-Tiny 4.1M
& $0.700 \pm 0.151$ 
& $0.806 \pm 0.160$ 
& $0.918 \pm 0.735$ \\
\hline

\end{tabular}
\vspace{-0.3cm}
\end{table}

\begin{table*}[!t]
\caption{Event-based performance of foundation-model adaptation strategies as mean $\pm$ standard deviation across LOO folds. 
Bold values indicate the best mean performance across models for each metric.}
\vspace{-0.4cm}
\label{tab:fm_adaptation_results}
\begin{center}
\scriptsize
\renewcommand{\arraystretch}{1.25}
\setlength{\tabcolsep}{4pt}

\begin{tabular}{|p{0.16\textwidth}|cc|cc|cc|}
\cline{2-7}
\multicolumn{1}{c|}{} 
& \multicolumn{2}{c|}{\textbf{LUNA}} 
& \multicolumn{2}{c|}{\textbf{REVE}} 
& \multicolumn{2}{c|}{\textbf{LuMamba}} \\
\hline

\textbf{Adaptation strategies} 
& \textbf{\textit{Event-based F1}} 
& \textbf{\textit{\begin{tabular}{c}Burst-per-minute MAE\end{tabular}}} 
& \textbf{\textit{Event-based F1}} 
& \textbf{\textit{\begin{tabular}{c}Burst-per-minute MAE\end{tabular}}} 
& \textbf{\textit{Event-based F1}} 
& \textbf{\textit{\begin{tabular}{c}Burst-per-minute MAE\end{tabular}}} \\
\hline

Fine-tuning 
& $0.838 \pm 0.120$ 
& $0.689 \pm 0.547$ 
& $\mathbf{0.868 \pm 0.167}$ 
& $\mathbf{0.448 \pm 0.284}$ 
& $0.806 \pm 0.160$ 
& $0.918 \pm 0.735$ \\
\hline

\begin{tabular}[c]{@{}l@{}}Fine-tuning (random weights)\end{tabular}
& $0.836 \pm 0.128$ 
& $0.743 \pm 0.559$ 
& $0.831 \pm 0.227$ 
& $0.513 \pm 0.335$ 
& $0.828 \pm 0.134$ 
& $0.799 \pm 0.589$ \\
\hline

Frozen backbone 
& $0.736 \pm 0.227$ 
& $0.899 \pm 0.722$ 
& $0.791 \pm 0.300$ 
& $0.550 \pm 0.432$ 
& $0.777 \pm 0.167$ 
& $0.816 \pm 0.542$ \\
\hline

LoRA 
& $0.726 \pm 0.232$ 
& $0.966 \pm 0.760$ 
& $0.753 \pm 0.328$ 
& $0.597 \pm 0.500$ 
& $0.776 \pm 0.166$ 
& $0.830 \pm 0.545$ \\
\hline

2-steps 
& $0.735 \pm 0.220$ 
& $0.906 \pm 0.721$ 
& $0.788 \pm 0.301$ 
& $0.553 \pm 0.440$ 
& $0.778 \pm 0.170$ 
& $0.811 \pm 0.539$ \\
\hline
\end{tabular}

\end{center}
\vspace{-0.6cm}
\end{table*}

In this study, we evaluated EEG FMs for burst detection in reduced-montage ICU EEG without patient-specific calibration. 
A key methodological contribution of this work is the use of the oriented event-based evaluation together with BPM error, providing a clinically interpretable assessment since BPM is regularly used for sedation titration. Window-based F1 evaluates local burst-versus-suppression classification on each EEG window, but it may penalize small onset or offset errors that have limited clinical relevance and may fall within annotation variability \cite{brandonwestoverRealtimeSegmentationBurst2013a}. This is particularly relevant when event boundaries are uncertain, as in marker-based annotations, or in tasks with high inter-rater variability. The lower window-based F1-scores in Table~\ref{tab:usz_results} support the need to evaluate automatic burst detection beyond conventional window-level metrics. Therefore, we focus the following discussion primarily on event-based F1 and BPM MAE, which better reflect clinically meaningful burst detection and burst-count estimation.

With regard to the performance comparison with the state-of-the-art, RVT remained a good baseline, with event-based F1 ($0.812 \pm 0.233$) close to LUNA-large ($0.838 \pm 0.120$). However, the best performing EEG FMs, REVE-base, achieved the highest event-based F1-score ($0.868 \pm 0.167$) and the lowest BPM MAE ($0.448 \pm 0.284$ burst/ min). This corresponds to a \(36.2\%\) reduction in BPM MAE compared with RVT, indicating more accurate and clinically interpretable burst-count estimation. At the same time, all finetuned FMs outperformed EEGNet, with REVE-base reducing BPM MAE by \(52.1\%\), suggesting that large pretrained architectures may be better suited than a convolutional model when trained from scratch on limited annotated data. Moreover, both REVE-base and LUNA-large achieved lower BPM MAE and substantially lower variability than the patient-specific unsupervised spectral-clustering method previously tested on the same USZ dataset (MAE = $0.93 \pm 1.38$ bursts/min)\cite{narulaDetectionEEGBurstsuppression2021a}.

Next, we investigate how the adaptation strategy and pretraining affected FM transfer to BS detection. As shown in Table~\ref{tab:fm_adaptation_results}, full fine-tuning was the most effective strategy, with improved event-based F1-score over frozen-backbone training by \(0.102\), \(0.077\), and \(0.029\) absolute F1 points for LUNA-large, REVE-base, and LuMamba-Tiny, respectively. Although two-step fine-tuning was intended to stabilize adaptation by first warming up the classification head, and LoRA was expected to provide a parameter-efficient way to adapt the encoder while limiting disruption of pretrained representations, both strategies underperformed full fine-tuning for all FMs. Overall, these results indicate that, in this setting, full encoder adaptation was more effective than either keeping the backbone fixed or applying limited low-rank updates. 

Because the evaluated FMs differ substantially in size, their direct comparison should not be interpreted as an isolated effect of model architecture. The stronger performance of REVE-base (69.2M) and LUNA-large (40.5M) compared with LuMamba-Tiny (4.1M) is consistent with a possible benefit of larger model capacity, but this effect remains confounded by other factors, like differences in pretraining settings and data.

\begin{figure}[!b]
    \centering
    \vspace{-0.5cm}\includegraphics[width=0.8\columnwidth]{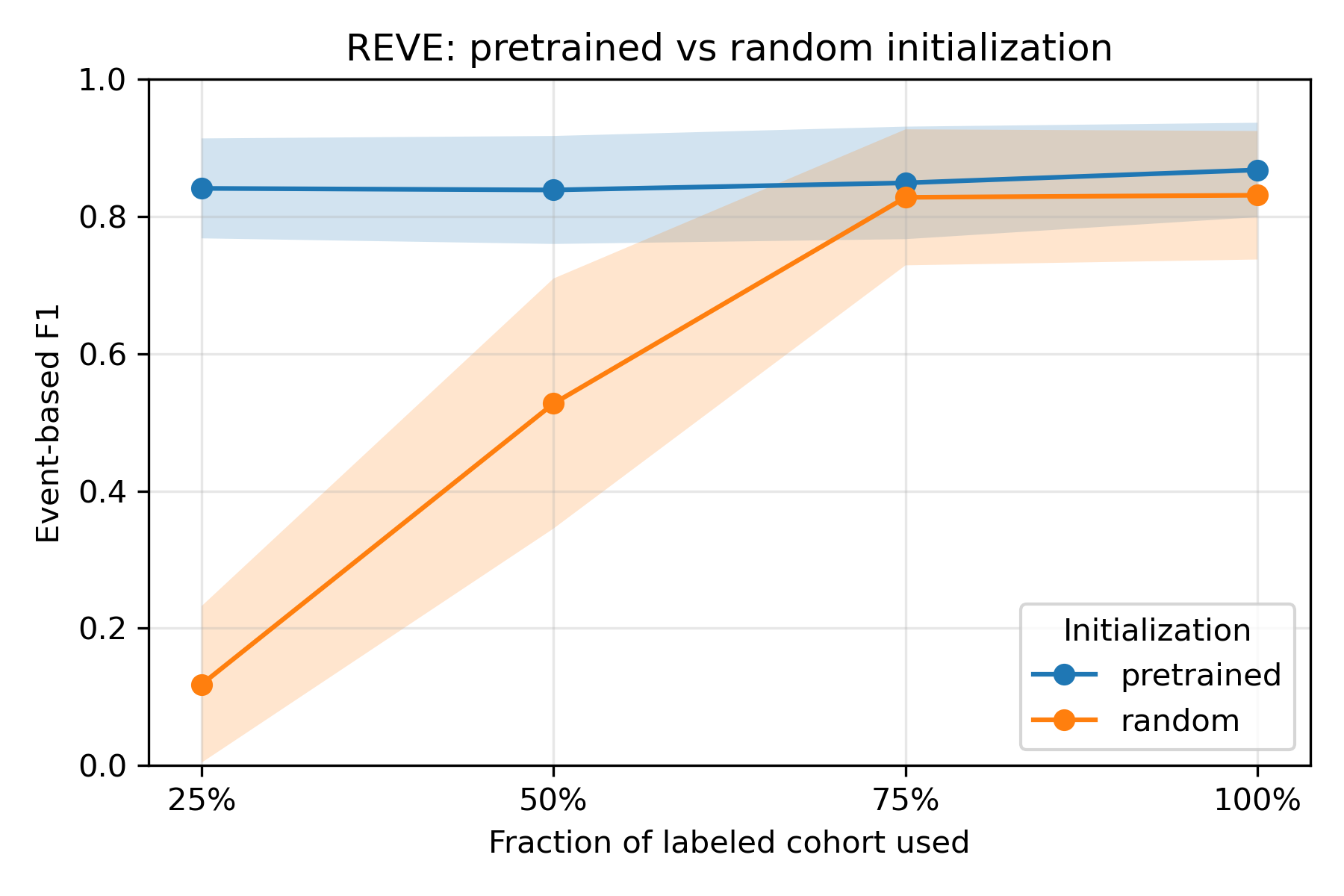}
    \vspace{-0.3cm}
    \caption{REVE-base performance (mean event-based F1-score ± 95\% CI) with pretrained versus random initialization across increasing cohort fractions.}
    \label{fig:subset_experiments}
\end{figure}

To assess the contribution of pretraining under limited labeled data, we compared fine-tuning from pretrained and random initialization while progressively increasing the fraction of available training data. In the full-cohort setting with around 21h of training data in each fold, REVE-base improved event-based F1 by \(+0.037\) and reduced BPM MAE by \(12.7\%\) compared with random initialization, whereas LUNA-large and LuMamba-Tiny showed only marginal F1-score differences. However, the benefit of pretraining became clearer when labeled training data were reduced.
As shown in Fig.~\ref{fig:subset_experiments}, random initialization remained competitive for REVE-base only when at least 75\% of the training data were available, corresponding to approximately 15 h 45 min. With only 25\% of training data, around 5 h 15 min, pretraining improved event-based F1-score by \(+0.723\) compared with random initialization. These results suggest that the main advantage of pretraining is improved data efficiency, enabling robust burst detection with substantially less annotated BS data.
The magnitude of this benefit was model-dependent, potentially reflecting differences in pretraining scale: REVE was pretrained on more than 60k hours of EEG from 92 datasets and 25k subjects ~\cite{ouahidiREVEFoundationModel}, whereas LUNA and LuMamba were pretrained only on 21k hours from TUEG and Siena~\cite{donerLUNAEfficientTopologyAgnostic2025}.

\section{Conclusion}
This study provides the first evaluation of EEG FMs for BS detection in ICU EEG. Fine-tuned FMs, especially REVE-base, supported generalized burst detection without patient-specific calibration, outperforming EEGNet and RVT by reducing BPM error by \(52.1\%\) and \(36.2\%\), respectively. 
The thresholding-based method remains a strong baseline because of its simplicity and competitive performance, but large pretrained FMs provide more accurate burst-count estimation under signal variability and artifacts.
Our results suggest that full fine-tuning of pretrained EEG FMs is the most effective strategy, and that pretrained representations are particularly useful when annotated BS data are scarce.
Finally, our results highlight the importance of evaluating BS detection beyond window-level classification. Event-based F1 should be used to assess burst-event detection performance, while BPM error should be reported when the intended use is clinical burst counting for sedation monitoring. 

Our study is limited with regard to the cohort size. Future work will focus on validating these findings on larger and possibly external cohorts, and investigate whether continued pretraining on ICU BS-specific recordings improves downstream adaptation.
 
\section*{Acknowledgment}
This study is part of the project ``SonNEEM: Sonified neurological EEG events monitoring" funded by the Digitalization Initiative of the Zurich Higher Education Institutions (DIZH).

\bibliographystyle{IEEEtran}
\bibliography{references}

@IEEEtranBSTCTL{IEEEexample:BSTcontrol,
  CTLuse_forced_etal       = "yes",
  CTLmax_names_forced_etal = "1",
  CTLnames_show_etal       = "1"
}

@article{alarcon-bragaDetectingDepthSedation2026,
  title = {Detecting the Depth of Sedation in the Intensive Care Unit Using a 2-Channel Electroencephalogram: {{An}} Analysis with 2 Machine Learning Models},
  shorttitle = {Detecting the Depth of Sedation in the Intensive Care Unit Using a 2-Channel Electroencephalogram},
  author = {{Alarc{\'o}n-Braga}, Esteban A. and Gruffaz, Samuel and Delagarde, C{\'e}cile and Roques, Axel and Riff, Jean-Cl{\'e}ment and Oudre, Laurent and Dubost, Cl{\'e}ment},
  year = 2026,
  journal = {Neuroscience Informatics},
  volume = {6},
  pages = {100238},
  doi = {10.1016/j.neuri.2025.100238},
  urldate = {2026-01-30},
  langid = {english}
}

@article{baldassanoIRISModularPlatform2020,
  title = {{{IRIS}}: {{A Modular Platform}} for {{Continuous Monitoring}} and {{Caretaker Notification}} in the {{Intensive Care Unit}}},
  shorttitle = {{{IRIS}}},
  author = {Baldassano, Steven N. and Roberson, Shawniqua Williams and Balu, Ramani and Scheid, Brittany and Bernabei, John M. and Pathmanathan, Jay and Oommen, Brian and Leri, Damien and Echauz, Javier and Gelfand, Michael and Bhalla, Paulomi Kadakia and Hill, Chloe E. and Christini, Amanda and Wagenaar, Joost B. and Litt, Brian},
  year = 2020,
  journal = {IEEE Journal of Biomedical and Health Informatics},
  volume = {24},
  doi = {10.1109/JBHI.2020.2965858},
  urldate = {2026-03-19}
}

@article{brandonwestoverRealtimeSegmentationBurst2013a,
  title = {Real-Time Segmentation of Burst Suppression Patterns in Critical Care {{EEG}} Monitoring},
  author = {Brandon Westover, M. and Shafi, Mouhsin M. and Ching, ShiNung and Chemali, Jessica J. and Purdon, Patrick L. and Cash, Sydney S. and Brown, Emery N.},
  year = 2013,
  journal = {Journal of Neuroscience Methods},
  volume = {219},
  pages = {131--141},
  doi = {10.1016/j.jneumeth.2013.07.003},
  urldate = {2025-09-22}
}

@misc{broustailLuMambaLatentUnified2026,
  title = {{{LuMamba}}: {{Latent Unified Mamba}} for {{Electrode Topology-Invariant}} and {{Efficient EEG Modeling}}},
  shorttitle = {{{LuMamba}}},
  author = {Broustail, Dana{\'e} and Tegon, Anna and Ingolfsson, Thorir Mar and Li, Yawei and Benini, Luca},
  year = 2026,
  eprint = {2603.19100},
  primaryclass = {cs},
  publisher = {arXiv},
  doi = {10.48550/arXiv.2603.19100},
  urldate = {2026-03-26},
  archiveprefix = {arXiv},
  langid = {english}
}

@article{chemaliBurstSuppressionProbability2013,
  title = {Burst Suppression Probability Algorithms: State-Space Methods for Tracking {{EEG}} Burst Suppression},
  shorttitle = {Burst Suppression Probability Algorithms},
  author = {Chemali, Jessica and Ching, ShiNung and Purdon, Patrick L and Solt, Ken and Brown, Emery N},
  year = 2013,
  journal = {Journal of Neural Engineering},
  volume = {10},
  pages = {056017},
  doi = {10.1088/1741-2560/10/5/056017},
  urldate = {2026-02-03},
  copyright = {http://iopscience.iop.org/info/page/text-and-data-mining},
  langid = {english}
}

@misc{donerLUNAEfficientTopologyAgnostic2025,
  title = {{{LUNA}}: {{Efficient}} and {{Topology-Agnostic Foundation Model}} for {{EEG Signal Analysis}}},
  shorttitle = {{{LUNA}}},
  author = {D{\"o}ner, Berkay and Ingolfsson, Thorir Mar and Benini, Luca and Li, Yawei},
  year = 2025,
  publisher = {arXiv},
  doi = {10.48550/ARXIV.2510.22257},
  urldate = {2026-02-10},
  copyright = {arXiv.org perpetual, non-exclusive license},
  langid = {english}
}

@article{furbassMonitoringBurstSuppression2016,
  title = {Monitoring Burst Suppression in Critically Ill Patients: {{Multi-centric}} Evaluation of a Novel Method},
  shorttitle = {Monitoring Burst Suppression in Critically Ill Patients},
  author = {F{\"u}rbass, Franz and Herta, Johannes and Koren, Johannes and Westover, M. Brandon and Hartmann, Manfred M. and Gruber, Andreas and Baumgartner, Christoph and Kluge, Tilmann},
  year = 2016,
  journal = {Clinical Neurophysiology},
  volume = {127},
  pages = {2038--2046},
  doi = {10.1016/j.clinph.2016.02.001},
  urldate = {2025-09-22}
}

@article{hirschAmericanClinicalNeurophysiology2021a,
  title = {American {{Clinical Neurophysiology Society}}'s {{Standardized Critical Care EEG Terminology}}: 2021 {{Version}}},
  shorttitle = {American {{Clinical Neurophysiology Society}}'s {{Standardized Critical Care EEG Terminology}}},
  author = {Hirsch, Lawrence J. and Fong, Michael W.K. and Leitinger, Markus and LaRoche, Suzette M. and Beniczky, Sandor and Abend, Nicholas S. and Lee, Jong Woo and Wusthoff, Courtney J. and Hahn, Cecil D. and Westover, M. Brandon and Gerard, Elizabeth E. and Herman, Susan T. and Haider, Hiba Arif and Osman, Gamaleldin and {Rodriguez-Ruiz}, Andres and Maciel, Carolina B. and Gilmore, Emily J. and Fernandez, Andres and Rosenthal, Eric S. and Claassen, Jan and Husain, Aatif M. and Yoo, Ji Yeoun and So, Elson L. and Kaplan, Peter W. and Nuwer, Marc R. and Van Putten, Michel and Sutter, Raoul and Drislane, Frank W. and Trinka, Eugen and Gaspard, Nicolas},
  year = 2021,
  journal = {Journal of Clinical Neurophysiology},
  volume = {38},
  pages = {1--29},
  doi = {10.1097/WNP.0000000000000806},
  urldate = {2026-04-07},
  langid = {english}
}

@article{maElectroencephalographicBurstSuppressionPerioperative2022,
  title = {Electroencephalographic {{Burst-Suppression}}, {{Perioperative Neuroprotection}}, {{Postoperative Cognitive Function}}, and {{Mortality}}: {{A Focused Narrative Review}} of the {{Literature}}},
  shorttitle = {Electroencephalographic {{Burst-Suppression}}, {{Perioperative Neuroprotection}}, {{Postoperative Cognitive Function}}, and {{Mortality}}},
  author = {Ma, Kan and Bebawy, John F.},
  year = 2022,
  journal = {Anesthesia \& Analgesia},
  volume = {135},
  pages = {79},
  doi = {10.1213/ANE.0000000000005806},
  urldate = {2026-01-20},
  langid = {american}
}

@article{narulaDetectionEEGBurstsuppression2021a,
  title = {Detection of {{EEG}} Burst-Suppression in Neurocritical Care Patients Using an Unsupervised Machine Learning Algorithm},
  author = {Narula, G. and Haeberlin, M. and Balsiger, J. and Str{\"a}ssle, C. and Imbach, L. L. and Keller, E.},
  year = 2021,
  journal = {Clinical Neurophysiology},
  volume = {132},
  pages = {2485--2492},
  doi = {10.1016/j.clinph.2021.07.018},
  urldate = {2025-09-22}
}

@article{ouahidiREVEFoundationModel,
  title = {{{REVE}}: {{A Foundation Model}} for {{EEG Adapting}} to {{Any Setup}} with {{Large-Scale Pretraining}} on 25,000 {{Subjects}}},
  author = {Ouahidi, Yassine El and Lys, Jonathan and Th{\"o}lke, Philipp and Farrugia, Nicolas and Pasdeloup, Bastien and Gripon, Vincent and Jerbi, Karim and Lioi, Giulia},
  langid = {english}
}

@article{hermanConsensusStatementContinuous2015,
  title = {Consensus {{Statement}} on {{Continuous EEG}} in {{Critically Ill Adults}} and {{Children}}, {{Part I}}: {{Indications}}},
  author = {Herman, Susan T and Abend, Nicholas S and Bleck, Thomas P and Chapman, Kevin E and Drislane, Frank W and Emerson, Ronald G and LaRoche, Suzette M and Nuwer, Marc R and Simmons, Liberty A and Tsuchida, Tammy N and Hirsch, Lawrence J},
  year = 2015,
  volume = {32},
  langid = {english}
}

@article{cottenceauUseBispectralIndex2008,
  title = {The {{Use}} of {{Bispectral Index}} to {{Monitor Barbiturate Coma}} in {{Severely Brain-Injured Patients}} with {{Refractory Intracranial Hypertension}}},
  author = {Cottenceau, Vincent and Petit, Laurent and Masson, Fran{\c c}oise and Guehl, Dominique and Asselineau, Julien and Cochard, Jean-Fran{\c c}ois and Pinaquy, Catherine and Leger, Alain and Sztark, Fran{\c c}ois},
  year = 2008,
  journal = {Anesthesia \& Analgesia},
  volume = {107},
  pages = {1676},
  doi = {10.1213/ane.0b013e318184e9ab},
  urldate = {2026-06-10},
  langid = {american}
}

@article{devlinClinicalPracticeGuidelines2018,
  title = {Clinical {{Practice Guidelines}} for the {{Prevention}} and {{Management}} of {{Pain}}, {{Agitation}}/{{Sedation}}, {{Delirium}}, {{Immobility}}, and {{Sleep Disruption}} in {{Adult Patients}} in the {{ICU}}},
  author = {Devlin, John W. and Skrobik, Yoanna and G{\'e}linas, C{\'e}line and Needham, Dale M. and Slooter, Arjen J. C. and Pandharipande, Pratik P. and Watson, Paula L. and Weinhouse, Gerald L. and Nunnally, Mark E. and Rochwerg, Bram and Balas, Michele C. and {van den Boogaard}, Mark and Bosma, Karen J. and Brummel, Nathaniel E. and Chanques, Gerald and Denehy, Linda and Drouot, Xavier and Fraser, Gilles L. and Harris, Jocelyn E. and Joffe, Aaron M. and Kho, Michelle E. and Kress, John P. and Lanphere, Julie A. and McKinley, Sharon and Neufeld, Karin J. and Pisani, Margaret A. and Payen, Jean-Francois and Pun, Brenda T. and Puntillo, Kathleen A. and Riker, Richard R. and Robinson, Bryce R. H. and Shehabi, Yahya and Szumita, Paul M. and Winkelman, Chris and Centofanti, John E. and Price, Carrie and Nikayin, Sina and Misak, Cheryl J. and Flood, Pamela D. and Kiedrowski, Ken and Alhazzani, Waleed},
  year = 2018,
  journal = {Critical Care Medicine},
  volume = {46},
  pages = {e825},
  doi = {10.1097/CCM.0000000000003299},
  urldate = {2026-06-10},
  langid = {american}
}

@article{loboDoesElectroencephalographicBurst2021a,
  title = {Does Electroencephalographic Burst Suppression Still Play a Role in the Perioperative Setting?},
  author = {Lobo, Francisco Almeida and Vacas, Susana and Rossetti, Andrea O. and Robba, Chiara and Taccone, Fabio Silvio},
  year = 2021,
  journal = {Best Practice \& Research Clinical Anaesthesiology},
  volume = {35},
  pages = {159--169},
  doi = {10.1016/j.bpa.2020.10.007},
  urldate = {2026-01-20},
  langid = {english}
}

@article{lofhedeClassificationBurstSuppression2008,
  title = {Classification of Burst and Suppression in the Neonatal Electroencephalogram},
  author = {L{\"o}fhede, J and L{\"o}fgren, N and Thordstein, M and Flisberg, A and Kjellmer, I and Lindecrantz, K},
  year = 2008,
  journal = {Journal of Neural Engineering},
  volume = {5},
  doi = {10.1088/1741-2560/5/4/005},
  urldate = {2026-05-26},
  langid = {english}
}

@inproceedings{orgucTimeFrequencyDomainClassifier2025,
  title = {Time-{{Frequency Domain Classifier}} for {{Propofol-Mediated Unconsciousness}}},
  booktitle = {2025 47th {{Annual International Conference}} of the {{IEEE Engineering}} in {{Medicine}} and {{Biology Society}} ({{EMBC}})},
  author = {Orguc, Sirma and Ardic, Fazil O. and Brown, Emery N.},
  year = 2025,
  pages = {1--7},
  doi = {10.1109/EMBC58623.2025.11254879},
  urldate = {2026-05-27}
}

@article{lawhernEEGNetCompactConvolutional2018,
  title = {{{EEGNet}}: A Compact Convolutional Neural Network for {{EEG-based}} Brain--Computer Interfaces},
  shorttitle = {{{EEGNet}}},
  author = {Lawhern, Vernon J and Solon, Amelia J and Waytowich, Nicholas R and Gordon, Stephen M and Hung, Chou P and Lance, Brent J},
  year = 2018,
  journal = {Journal of Neural Engineering},
  volume = {15},
  pages = {056013},
  doi = {10.1088/1741-2552/aace8c},
  urldate = {2026-02-11},
  langid = {english}
}

@misc{huLoRALowRankAdaptation2021a,
  title = {{{LoRA}}: {{Low-Rank Adaptation}} of {{Large Language Models}}},
  shorttitle = {{{LoRA}}},
  author = {Hu, Edward J. and Shen, Yelong and Wallis, Phillip and {Allen-Zhu}, Zeyuan and Li, Yuanzhi and Wang, Shean and Wang, Lu and Chen, Weizhu},
  year = 2021,
  eprint = {2106.09685},
  primaryclass = {cs.CL},
  publisher = {arXiv},
  doi = {10.48550/arXiv.2106.09685},
  urldate = {2026-06-11},
  archiveprefix = {arXiv}
}

@article{danSzCORESeizureCommunity2025,
  title = {{{SzCORE}}: {{Seizure Community Open-Source Research Evaluation}} Framework for the Validation of Electroencephalography-Based Automated Seizure Detection Algorithms},
  shorttitle = {{{SzCORE}}},
  author = {Dan, Jonathan and Pale, Una and Amirshahi, Alireza and Cappelletti, William and Ingolfsson, Thorir Mar and Wang, Xiaying and Cossettini, Andrea and Bernini, Adriano and Benini, Luca and Beniczky, S{\'a}ndor and Atienza, David and Ryvlin, Philippe},
  year = 2025,
  journal = {Epilepsia},
  volume = {66},
  pages = {14--24},
  doi = {10.1111/epi.18113},
  urldate = {2026-06-11},
  langid = {english}
}

\end{document}